\documentclass[12pt,preprintnumbers]{article}
\usepackage{amsmath, amssymb, wasysym, slashed, multirow,hyperref}
\usepackage{pdfpages}
\usepackage{cite}
\usepackage{dsfont}
\usepackage{times}
\usepackage{xcolor}
\usepackage{graphicx,color}  %
\usepackage{bm}  %

\newenvironment{sciabstract}{%
\begin{quote} }
{\end{quote}}

\newcounter{lastnote}

\usepackage{fancyhdr}
\fancypagestyle{plain}{%
	\fancyhead[R]{}

}

\title{A Consistent Quantum Field Theory from Dimensional Reduction}

\author
{Alessio Maiezza,$^{1\ast}$  Juan Carlos Vasquez$^{2\dagger}$\\
\\
\normalsize{$^{1}$Dipartimento di Scienze Fisiche e Chimiche, Universit\`a} \\ \\
\normalsize{degli Studi dell'Aquila, via Vetoio, I-67100, L'Aquila, Italy,}\\ \\
\normalsize{$^{2}$Department of Physics $\&$ Astronomy, Amherst College, Amherst, MA 01002, USA} \\
\\
\small{ E-mail: alessiomaiezza@gmail.com$^{\ast}$, jvasquezcarmona@amherst.edu$^{\dagger}$}
}

\date{}

\begin{document}


\baselineskip16pt 


\maketitle


\begin{sciabstract}
We incorporate the concept of dimensional reduction at high energies within the perturbative formulation of quantum field theory. In this new framework, space and momentum integrations are modified by a weighting function incorporating an \emph{effective} mass energy associated with the dimensional reduction scale. We quantize the theory within canonical formalism. We then show that it can be made finite in perturbation theory, free of renormalon ambiguities, and with better analytic behavior for infinitesimal coupling constant compared to standard quantum field theory. The new approach reproduces the known results at low energies. One \emph{key feature} of this class of models is that the coupling constant always reaches a fixed point in the ultraviolet region, making the models ultra-violet complete.
\end{sciabstract}

\section{Introduction\label{Introduction}}

The central object in quantum field theory (QFT) is the S-matrix, which is at the core of scattering evaluations and connects the quantization formalism to physical observables. While there are rigorous, non-perturbative definitions of the S-matrix --  see, e.g., Refs.~\cite{Haag1964,PhysRev.112.669,ruelle1962asymptotic} --  these \emph{particular} constructions lack contact with any fundamental theory, such as quantum electrodynamics (QED) or the standard model (SM). Today, the Lagrangian field theories need to be treated using perturbation theory; one quantizes the free field(s) and then evaluates the S-matrix elements using the Feynman rules. In doing this, one ignores that within the interaction picture,  this particular construction of the S-matrix is prevented by Haag's theorem (HT)~\cite{Haag:1955ev}. Not surprisingly, calculations in perturbation theory of non-trivial S-matrix elements give divergent results, which are cured by the renormalization procedure for renormalizable models. By subtracting these infinities, renormalization \textit{a posteriori} forces the S-matrix to exist. In other words, by discriminating between the initial (infinite) bare parameters and the renormalized (finite) ones, renormalization \emph{de facto} makes the non-interactive and interactive theories non-unitary equivalent. The free and the interacting models are unitary equivalent if there exists some \emph{a priori, finite} unitary matrix (Dyson matrix) connecting them, and this is one of the assumptions of HT. Dyson matrix is infinite in perturbation theory before renormalization, and hence not suitable to connect the free and interactive theories. Therefore, renormalization circumvents HT by adding some external information designed to make the theory finite, while unitary equivalence between the free and interacting theory is effectively lost.

Unlike other areas of physics, perturbation theory in QFT is not only an approximation technique but also an integral part of the renormalization procedure. Although it is a consistent procedure -- order-by-order -- an improved perturbation theory should be unitary equivalent to the free theory. Once one abandons the requirement of unitary equivalence, it is not guaranteed that the procedure obtained is complete, and this is what happens in the standard QFT. The renormalized asymptotic expansions in the coupling constant need to be resummed to achieve consistent results, but there are strong indications that such resummation is unattainable in four dimensions. Indeed, within $\phi^4$ model or QED, a non-ambiguous resummation is prevented by the ultra-violet (UV) renormalons, which shows the limitations of the renormalization program~\cite{tHooft:1977xjm}. In addition, there is the issue of the ``horned-shaped" analyticity domain in the complex coupling constant plane, which implies that the Borel transform grows faster than any exponential in the Borel variable, preventing the Borel-Laplace resummability of any four-dimensional (4D) QFT, even for asymptotically free models~\cite{tHooft:1977xjm}. The latter is also consistent with a class of diagrams, found in Ref.~\cite{deCalan:1981szv}, which makes the Laplace integral divergent.   
It is worth stressing that these renormalization issues are absent for super-renormalizable models. For example, the $\phi^4$ model in two dimensions is Borel resummable~\cite{glimm2012quantum}. Hence, for superrenormalizable models, perturbative renormalization suffices to obtain complete results. The reason is that only a finite set of graphs is divergent for super-renormalizable models. Thus less information has to be added for the consistency of the interactive theory.

In this work, we propose dimensional reduction at high energy as a solution to the problems mentioned in the standard, 4D QFT. There are indications from quantum approaches to gravity that space-time dimension might reduce at high-energy~\cite{tHooft:1993dmi,Lauscher:2005qz,Ambjorn:2005db,Benedetti:2008gu,Calcagni:2009kc,Calcagni:2010bj,Shirkov:2010sh,Modesto:2011kw}, and this possibility has been already considered in light of the current particle phenomenology~\cite{Anchordoqui:2010er} and astronomy~\cite{Mureika:2011bv}.
Focusing on the $\phi^4$ model for simplicity, we formulate a QFT with an energy-dependent space-time dimension and develop a canonical quantization formalism consistent with dimensional reduction. Within this scenario, we introduce a mass-energy scale that signals the change in the space-time dimension. The new scale can be thought of as the remnant of an unknown UV dynamics of space-time, not in the usual Wilsonian sense, i.e. not through higher-order operators suppressed by the scale of heavy particles, but instead through a \emph{classical} field of geometrical origin that weights differently long and short distances contributions~\footnote{The idea that very high energy dynamics might affect low energy physics has been known in the literature. For instance, in Ref.~\cite{Ellis:1992eh}, it was argued that UV stringy dynamics might modify quantum mechanics.}.
We call the resulting theory ``dimensionally-reduced QFT" (DRQFT) and show that, while keeping unitary equivalence between the free interacting theories, it avoids the problems of the standard QFT mentioned above. As we shall show, in DRQFT, the vacuum state would not be translation invariant; hence, it avoids one of the other assumptions leading to HT. As a result,  computations in perturbation theory within DRQFT can be made finite, with no renormalon singularities, and no bad analytic properties for infinitesimal coupling.

The structure of this article is as follows. In Sec.~\ref{sec:motivation}, we motivate the possibility of an energy scale in QFT associated with the space-time integration measure. The main results are discussed in Secs.~\ref{sec:quantization} and~\ref{SEC:loops}, in which we perform the canonical quantization and compute one-loop examples within the new theory. We also study the running coupling in the $\phi^4$ model and show the absence of renormalons and the good analytic properties of the theory for infinitesimal couplings. In Sec.~\ref{Discussion}, we present an outlook and discuss possible implications for realistic models. The paper is complemented by two appendices~\ref{SEC:Haag} and~\ref{sub:evasion}, in which we elaborate on the  HT and its implications for perturbation theory.
Finally, we suggest a way of avoiding the no-go imposed by HT.

\section{Dimensionally-reduced quantum field theory}\label{sec:motivation}

In this section, we discuss the basis needed to elaborate on the DRQFT and discuss how our proposal relates to the current literature. In Refs.~\cite{Calcagni:2009kc,Calcagni:2010bj}, the author implements multifractal modifications of the physical dimensions that might also include fractional operators~\cite{Calcagni:2011sz}. Previous attempts to model QFT on fractal space-time can be found in Refs.~\cite{svozil1987quantum,eyink1989quantum}. The proposed DRQFT is not equivalent to the multifractal approach of Ref.~\cite{Calcagni:2010bj}, although both theories formally share the classical-field structure. In particular, in DRQFT, there is a new energy scale signaling the reduction of space-time.
This implies a different structure at the quantum level. In this sense, our approach is \emph{effective} and closer to the one of Ref.~\cite{Shirkov:2010sh}. There, the author provides a heuristic picture of dimensional reduction versus the running coupling, via an \emph{ansatz} for the momentum integration. An akin one can be derived from the canonical quantization formalism adapted to a scale-dependent dimensionality of space-time.

The resulting DRQFT is finite in perturbation theory, with no renormalons and likely Borel resummable~\footnote{There are other possible sources of ambiguity, the instanton\cite{Belavin:1975fg,tHooft:1976snw}. However, these have a semi-classical limit that, in principle, enables one to fix the ambiguities in the Laplace integral -- see, for example, Ref.~\cite{Coleman:1978ae,zinn2011barrier}. In this sense, the instantons do not damage the consistency of the perturbative QFT.}. Furthermore, DRQFT evades the no-go imposed by Haag's theorem, and we refer the reader to  Appendix~\ref{SEC:Haag} for more details.

\subsection{A new mass/energy scale}

In standard QFT, a renormalizable Lagrangian $\mathcal{L}$ is made effective by adding higher-dimensional operators $O_i$:
\begin{equation}\label{effective}
S= \int d^4 x \mathcal{L} \rightarrow \int d^4 x \left(\mathcal{L} + \sum \frac{O_i}{M^i} \right) \,,
\end{equation}
where the operators $O_i$ are suppressed by some ``new scale" $M$. The meaning of Eq.~\eqref{effective} is that these operators are obtained after integrating out some heavy particles with mass $M$.\\

It is also conceivable that the dimension of space-time is energy-dependent and that the new mass-scale $M$ is not associated with some new heavy particles -- as shown in Eq.~\eqref{effective} -- but with an intrinsic energy scale signaling the change in the space-time dimension. As a first approach, we consider a QFT embedded in a flat space-time, where the space-time dimension is energy-dependent, and gets smaller at high energies. The latter is motivated by the fact that most known approaches to gravity at the most fundamental level suggest that the space-time dimension is less than four at high energies -- see the reviews in Refs.~\cite{Stojkovic:2013xcj,Carlip:2017eud}.  Notice that there is no consensus about how the dynamical dimensional reduction should work. Moreover, the definition of physical dimension is inherently tricky at the microscopic level, and one can only appeal to different dimensional estimators~\cite{Carlip:2017eud}.

At energy  $E<<M$, we assume an effective dimension coinciding with the standard topological one (4D), while it \emph{effectively} reduces to a lower dimension at $E\sim M$. This approach has been suggested in Ref.~\cite{Shirkov:2010sh},  where the author performed a ``hard conjunction" between a 4D and a 2D Lagrangians at  $E=M$. Notice that the choice of four dimensions at low energies is empirical and the formalism discussed below is generalizable to an arbitrary topological dimension. 

We describe the smooth change in the space-time dimension in terms of Lebesgue-Stieltjes integration measures in coordinate (and momentum space) as
$d^4 x \mapsto d w(x)$ (and $d^4 k \mapsto d w(k)$). To confront the familiar dimensional regularization, one can regard the measures as
\begin{align}\label{w}
&  d w(x) := M^{-\alpha(x)} \, d^{D(x)} x \hspace{3em}\text{with}\hspace{3em} D(x):=4-\alpha(x)  \nonumber \\
&  d w(k) := M^{\alpha(k)} \, d^{D(k)} k  \hspace{3.65em}\text{with}\hspace{3em} D(k):=4-\alpha(k)      \,.
\end{align}
The function $\alpha(k)$ parameterizes the dependence of space-time dimension as a function of the energy and -- as we shall see in detail -- has to be small only at deep IR, to match with standard QFT (i.e. usual four-dimensional framework). Notice also that $w(x)$ and $w(k)$ assume the same functional form since probing short distances corresponds to probing high-energy scales, in agreement with Heisenberg's uncertainty principle.

 The Eq.~\eqref{w} resembles dimensional regularization but with a physical energy scale $M$ in place of the benchmark energy $\mu$ and an energy-dependent dimension $D(k)$. If $D(k)$ were constant, one would recover dimensional regularization. In this sense, within this approach, there is an actual physical change in the space-time dimension, unlike dimensional regularization. We should clarify that Eq.~\eqref{w} only serves to compare with the standard dimensional regularization, but it has no application for the rest of the paper.
Indeed, the momentum dependence of the space-time dimension entails an intrinsic difficulty in evaluating any integration. It can be overcome by properly handling all the integrals in the theory as Lebesgue-Stieltjes ones. Consistently, one must start with the action written as a Lebesgue-Stieltjes integral~\cite{Calcagni:2010bj}:
\begin{equation}\label{new_effective}
S=\int dw(x) \mathcal{L}  \,.
\end{equation}
Short and long distances are ``weighted" differently -- hence the name $w(x)$. Unlike Eq.~\eqref{effective}, Eq.~\eqref{new_effective} may also be seen as an effective action because it effectively describes the reduction of the space-time dimension at high energies. One can define the Lebesgue-Stieltjes measure in each space-time direction~\cite{Calcagni:2010bj}:
\begin{equation}\label{measure_expand}
dx =\prod_{i=0}^{3} dx_i  \mapsto dw(x):= \prod_{i=0}^{3} dx_i \,s_i(x)  \,,
\end{equation}
where we denote $dx =d^4 x$ (we shall also denote $d\bar{x}=d^3 x$, and similarly for the momentum space).
We shall assume
that the weight $s_0(x) =1 $ on the temporal direction,  and $s_i=s(x)$ with $i=1,2,3$. With these assumptions, the integration measure can be written as
\begin{equation}\label{measure}
 dw(x) :=  dx \, r(x)= dx_0 \, d\bar{x} \, r(x)\,,
\end{equation}
with $r(x)=s(x)^3$.

In contrast to Ref.~\cite{Calcagni:2010bj}, from Eq.~\eqref{w}, it follows that the function $r$ is dimensionless -- then a function of  $x\times M$ or $k/M$ in coordinate and momentum space, respectively. In our case, the dimension of the Lagrangian is as in standard QFT. The objectives of Ref.~\cite{Calcagni:2010bj} are different from ours, namely, the author attempts a formulation of perturbatively renormalizable quantum gravity.

Similarly to Eq.~\eqref{measure}, one has from Eq.~\eqref{w} the weight in momentum space
\begin{equation}
dk \mapsto dw(k) =  dk \, r(k)= dk_0\, d\bar{k} \, r(k) \,.
\end{equation}
having $r(x)$ and $r(k)$ the same functional form.\\

Before discussing further technical issues, a comment is in order. At first glance, Eq.~\eqref{new_effective} resembles string theory dilaton, in which
the field $r(x)$ can be thought of a dilaton, typically defined as $r(x)\approx e^{-\Phi(x)}$~\cite{Lovelace:1983yv,Fradkin:1984pq,Fradkin:1985fq,Callan:1985ia}. There are, however, deep distinctions between the dilaton models and dimensional reduction cases. Albeit both the dilaton and the above field $r$ are of geometrical origin, the dilaton couples differently, for different sectors of a quantum field theory, while
$r$ is a global rescaling that manifests itself through a change of the integration measure, together with a re-definition of the calculus -- see next paragraph. This is a central point for the rest of the paper. The modified calculus plays a crucial role in defining a Dirac-like distribution which, in turn, is a fundamental object for the canonical quantization that we want to perform.

\paragraph{Calculus.}
The introduction of the Lebesgue-Stieltjes integration requires some specific definitions in calculus. A generalized delta function is defined as~\cite{svozil1987quantum}
\begin{equation}\label{d1}
\int dw(x) \, \delta(x) = 1\,,
\end{equation}
and similarly in momentum space. We denote the four-dimensional delta function as  $\delta^{(4)} (x) =\delta(x)$. Conversely, we shall denote explicitly with $\delta^{(1)}$ and $\delta^{(3)}$ the one-dimensional (function of the temporal variable) and three-dimensional (function of the space variables), respectively. One has also to define the Lebesgue-Stieltjes-Fourier representation for the delta function,
\begin{equation}\label{d2}
\int dw(x) \, e^{i (k-k')x}=(2\pi)^4\delta(k-k')\,,
\end{equation}
and
\begin{equation}\label{d3}
\int dw(k) \, e^{-i (x-x')k}=(2\pi)^4\delta(x-x')\,.
\end{equation}
The Lebesgue-Stieltjes-Fourier transform for the classical field is
\begin{align}
\tilde{\phi}(k) &= \int dw(x) \, e^{i k x}\phi(x)\,,  \label{Fouriernew1} \\
\phi(x) &= \frac{1}{(2\pi)^4}\int dw(k) \,  e^{-i k x} \tilde{\phi}(k)\,.  \label{Fouriernew2}
\end{align}
such that, replacing the  Eq.~\eqref{Fouriernew1} in  Eq.~\eqref{Fouriernew2}  one obtains Eq.~\eqref{d2}. In this way, the set of Eqs.~\eqref{d1}--~\eqref{Fouriernew2} generalizes the standard calculus.

\subsection{Tree-level scale invariance and the Callan-Symanzik equation}

We show that the scale invariance of the action functional is not spoiled in the DRQFT. To this end, consider the action functional
\begin{equation}\label{action}
S=\int dw(x) \mathcal{L} = \int dx \, r(x) \mathcal{L}\,,
\end{equation}
being $\mathcal{L}= \frac{1}{2} \partial_\mu \phi \partial^\mu \phi$ the Lagrangian density of a free, massless, real scalar field $\phi$. An \emph{infinitesimal} scale transformation on the scalar field is given by
\begin{equation}
 \delta\phi(x)=\phi'(x')-\phi(x) = (d+x^{\mu}\partial_{\mu})\phi(x)\,,
\end{equation}
where  $d=1$ is the so-called scale dimension for the scalar field. Under a scale transformation, the action varies as
\begin{equation}
\delta S  =  \int dx\, \delta r(x) \, \mathcal{L}(x) +  \int dx\,  r(x) \,\delta \mathcal{L}(x) \,.
\end{equation}
By construction, the Lagrangian is a scalar with dimension four, then it transforms as $\delta \mathcal{L}(x) = (4+ x^{\mu}\partial_{\mu}) \mathcal{L}(x) $. The variation of the action can be written as
\begin{align}
\delta S & =  \int dx\, \delta r(x) \, \mathcal{L}(x) +  \int dx\,  r(x) \,(4+ x^{\mu}\partial_{\mu}) \mathcal{L}(x) \nonumber \\  
& = \int dx\, \delta r(x) \, \mathcal{L}(x)  + \int dx \,\partial_{\mu} \left(  r(x) \,x^{\mu} \mathcal{L}(x)	\right)
- \int dx \,  \mathcal{L}(x)\, x^{\mu}\partial_{\mu}r(x) \,,
\end{align}
in which we assume that the total derivative term vanishes. As usual, this can be achieved by assuming that the Lagrangian and the fields vanish faster than $1/|x|$ at infinity. The same assumption shall be made when deriving the equations of motion.

 If one assumes that $r(x)$ is a dimensionless scalar field, then under scale transformations $\delta\,r(x) = x^{\mu}\partial_{\mu} r(x) $ and in this case $\delta S =0 $, i.e., scale invariance is preserved. Since scale invariance is not broken at the tree level, it implies that the $n$-point Green functions still satisfy the  Callan-Symanzik equation~\cite{PhysRevD.2.1541,Symanzik:1973pp}. Therefore, in momentum space, the Feynman propagator assumes the same form as in standard QFT. This is undoubtedly an asset of the theory because there shall be no substantial modifications in the well-known machinery of loop calculations. We should stress that this is a specific consequence of the assumption that the scale $M$ enters into $r(x)$  as a dimensionless ratio. As already discussed, this is not the case with the approach in Ref.~\cite{Calcagni:2010bj}, in which the weight function is dimensionful. However, this difference shall not affect the equation of motion, which we shall show in the following subsection following Ref.~\cite{Calcagni:2010bj}.

\subsection{Classical field theory}

Consider the free, massive scalar field with Lagrangian density.
\begin{equation}\label{freeL}
\mathcal{L}= \frac{1}{2} \left( \partial_\mu \phi \partial^\mu \phi -m^2 \phi^2 \right) \,,
\end{equation}
The minimization of the action in Eq.~\eqref{action} leads to the equation of motion
\begin{equation}\label{KGE}
\left( \Box+\frac{ \partial_\mu r(x) }{r(x)}\partial^\mu-m^2\right)\phi(x) = 0  \,,
\end{equation}
where $\Box := \partial_\mu \partial^\mu$. The energy-momentum tensor is given by
\begin{equation}
\mathcal{T}^{\mu\nu}:= \frac{\partial\mathcal{L}}{\partial(\partial_\mu\phi)}\partial^\nu \phi - \mathcal{L} g^{\mu\nu}\,,
\end{equation}
and we use  the convention for the metric $g^{\mu\nu}=\text{diag}\{1,-1,-1,-1\}$. Considering a translation of an infinitesimal parameter $b$, then $\delta\phi=-\partial^\mu \phi b_\mu$
and from Eq.~\eqref{action}
\begin{equation}
\delta S= -r(x) \partial_\mu \mathcal{L} b^\mu - \mathcal{L} \partial_\mu r(x) b^\mu\,,
\end{equation}
and then
\begin{equation}\label{continuity}
\partial_\mu (r(x) \mathcal{T}^{\mu\nu})= -\mathcal{L} \partial^\nu r(x)\,.
\end{equation}
Defining
\begin{equation}
P_\mu= \int d\bar{x} \, r(x) \mathcal{T}_\mu^0\,,
\end{equation}
Eq.~\eqref{continuity} implies
\begin{equation} \label{pdot}
\dot{P}_\mu= -\int d\bar{x} \, \partial_\mu r(x) \mathcal{L}\,,
\end{equation}
where the dot denotes the time derivative. For the scalar field, the time derivative of the 3-momentum is of the form
\begin{equation}\label{3momentum}
P_i= -\int d\bar{x} \, r(x) \dot{\phi} \partial_i \phi\,,
\end{equation}
where translation operator is given by $T=e^{-i P_i b^i}$. Notice that from Eq.~\eqref{pdot}, the operator $T=e^{-i \vec{P}\cdot \vec{b}}$ for space translations is time-dependent, and we refer the reader to the Appendix~\ref{sub:evasion} for more details and implications.

The bottom line is that the weight $r(x)$, which is nontrivial only at energy $\gtrsim M$, modifies the translation within the Poincar\'e group. Notice that the weight modifies also the Lorentz group generators, but the Lorentz algebra is preserved. We refer the reader to Ref.~\cite{Calcagni:2010bj} for a detailed discussion.

\section{Canonical quantization}\label{sec:quantization}

In this section, we perform the canonical quantization for the DRQFT. First, replacing the Stieltjes-Fourier transform of $\phi$ in Eq.~\eqref{Fouriernew2} into Eq.~\eqref{KGE}, one gets
\begin{equation}
\frac{1}{(2\pi)^4}\int dk \, r(k)\left(-k^2-i k_\mu \frac{\partial^\mu r(x)}{r(x)}+m^2\right) e^{-i k x} \tilde{\phi}(k)\,.
\end{equation}
This gives the following equations
\begin{align}
& k^2-m^2=0 \label{standard_dispersion} \\
& k_\mu \, \partial^\mu r(x)=0   \label{dispersion} \,.
\end{align}
The first equation is just the standard dispersion relation, while the second gives an additional constraint to the function $r(x)$. In particular, it implies that $r$ cannot be a function of only $x^2$: if this were the case, it is easy to see that taking the derivative of $r(x^2)$, the Eq.~\eqref{dispersion} would lead to $k_\mu x^\mu=0$. The latter would prevent a non-trivial Fourier representation since both momentum and position would not be independent variables. As a consequence, $x^{\mu}$ must appear at least linearly inside $r$, and we must introduce a four-vector parameter $a^\mu$ such that
\begin{equation}\label{constraint}
r=r(x_\mu a^\mu), \hspace{4em} |a|\sim 1/M \,,
\end{equation}
with $M$ the dimensional reduction scale of Sec.~\ref{sec:motivation}. Using Eq.~\eqref{dispersion}, one obtains $k_{\mu}a^{\mu} =0$, which is a restriction on the possible momenta for on-shell particles.  Since $|a|\sim 1/M$, the latter constraint becomes relevant only when the energy is of order or bigger than $M$.

\paragraph{Quantization and canonical commutation relation.} The canonical commutation relation (CCR) is
\begin{equation}\label{CCR}
[\phi(t,\bar{x}),\pi(t,\bar{y})]= i \delta^{(3)}(\bar{x}-\bar{y})\,,
\end{equation}
being $\pi=\frac{\partial\mathcal{L}}{\partial\dot{\phi}}=\dot{\phi}$, but with the difference that the delta function is the generalized version defined below Eq.~\eqref{d1}.  Following the standard procedure, we next introduce the ladder operators $a(\bar{k}),a^\dagger(\bar{k})$ to rewrite the Eq.~\eqref{Fouriernew2} as a quantum field:
\begin{equation}\label{quantumfield}
\phi(x)= \frac{1}{(2\pi)^3} \int \frac{d\bar{k}}{\sqrt{2 \omega_k}}  \left[  r(\bar{k}) a(\bar{k}) e^{-i k x} +  r(-\bar{k}) a^\dagger(\bar{k}) e^{i k x}\right]  \,,
\end{equation}
and thus
\begin{equation}\label{quantumfield2}
\pi(x)= -\frac{i}{(2\pi)^3} \int d\bar{k} \sqrt{\frac{\omega_k}{2}} \left[ r(\bar{k}) a(\bar{k}) e^{-i k x} -  r(-\bar{k}) a^\dagger(\bar{k}) e^{i k x}\right]\,.
\end{equation}
We are denoting $r(\bar{k})=r(k)|_{k_0=\omega_k}$ with $\omega_k=\sqrt{(\vec{k})^2+m^2}$, namely the usual dispersion in Eq.~\eqref{standard_dispersion}.

The Eq.~\eqref{quantumfield} is derived from the classical field with standard manipulations, and when one changes the second piece in the integral
$\bar{k}\rightarrow -\bar{k}$ also splits the function $r$ into two parts (with opposite signs in the argument) because $r$ is not a function of $\bar{k}^2$. This follows from Eq.~\eqref{constraint} (and the fact that the function $r$ has the same form both in the coordinate and momentum space, as discussed in Sec~\ref{sec:motivation}). The Eqs.~\eqref{quantumfield} and~\eqref{quantumfield2} give the commutation relation
\begin{equation}\label{CCR2}
[a(\bar{k}), a^\dagger(\bar{k'})] = \delta^{(3)}(\bar{k}-\bar{k'})\,,
\end{equation}
which, again, looks like the standard one, except that the delta is the generalized one.

\paragraph{3-momentum.}
Taking the space derivative of  Eq.~\eqref{quantumfield} gives
\begin{equation}\label{quantumfield3}
\partial_i\phi= \frac{-i}{(2\pi)^4} \int d\bar{k} \frac{\bar{k}_i}{\sqrt{2 \omega_k}}  \left[ r(\bar{k}) a(\bar{k}) e^{-i k x} - r(-\bar{k}) a^\dagger(\bar{k}) e^{i k x}\right]\,,
\end{equation}
which, once replaced in Eq.~\eqref{3momentum} and after some manipulations gives
\begin{equation}\label{quantumPi}
\int P_i \, dx_0 = \delta^{(1)}(0) \int d\bar{k} \, \frac{\bar{k}_i}{2} [ r(-\bar{k}) a^\dagger(\bar{k}) a(\bar{k}) + r(\bar{k}) a(\bar{k}) a^\dagger(\bar{k})]\,.
\end{equation}
being
\begin{equation}
\delta^{(1)}(0) :=   \lim_{\bar{k}'\rightarrow \bar{k}}\delta^{(1)}(\omega_k-\omega_{k'})
\end{equation}
a c-number, as in the standard case. Since momentum is not constant in time, the above equation can be interpreted as the time average of the 3-momentum.  Technically, this comes from the necessity to complete the measure $dx r(x)$, starting from Eq.~\eqref{3momentum} and using expression~\eqref{d2}.

\paragraph{Action of $P_i$ on the vacuum.} The Fock space $\mathcal{F(\mathcal{H})}=\bigoplus_{n=0}^\infty \mathcal{H}_0^{\bigotimes n}$ (understanding symmetrization for bosonic states of $\phi$) is spanned as in the standard case by the ladder operators. The $n-$particle states are normalized, as usual
\begin{equation}
\langle n_k | n_{k'} \rangle = \delta^{(3)}(\bar{k}-\bar{k}')\,,
\end{equation}
such that
\begin{equation}
\int d\bar{k}' \, r(\bar{k}') \langle n_k | n_{k'} \rangle = 1\,,
\end{equation}
which is the weighted orthonormality relation. Taking the vacuum-to-vacuum expectation value of Eq.~\eqref{quantumPi}, one obtains
\begin{equation}\label{3momentum_integrated}
\int \langle 0 | P_i | 0 \rangle \, dx_0 = \delta(0)^{(3)}\delta(0)^{(1)} \int d\bar{k} \, r(\bar{k}) \frac{\bar{k}_i}{2}=\delta(0) \int d\bar{k} \, r(\bar{k}) \frac{\bar{k}_i}{2}\,,
\end{equation}
where
\begin{equation}
\delta^{(3)}(0) :=   \lim_{\bar{k}'\rightarrow \bar{k}}\delta^{(3)}(\bar{k}-\bar{k}')\,.
\end{equation}
Notice that in the standard limit $r(x)\rightarrow 1$ and the Eq.~\eqref{3momentum_integrated}, being an odd function of $\bar{k}$, is equal to zero. In the DRQFT, Eq.~\eqref{3momentum_integrated} is non-zero since $r(\bar{k})\neq r(-\bar{k})$, because of Eq.~\eqref{constraint}. Therefore, contrary to the standard QFT, in DRQFT, the vacuum expectation value of the 3-momentum is time-dependent and given by
\begin{equation}
\langle 0 | P_i |0\rangle = \int d\bar{x}\,  r(x_0,\bar{x}) \int d\bar{k} \, r(\bar{k}) \frac{\bar{k}_i}{2}\,,
\end{equation}
which implies
\begin{equation}
  P_i |0\rangle\neq 0\,.
\end{equation}
Finally, using the expression
\begin{equation}
T_0=e^{i P_i b_i}=1+i P_i b_i+\mathcal{O}(b_i)^2 \,,
\end{equation}
it is easy to see that
\begin{equation}\label{achievement_evasion}
T_0| 0 \rangle \neq | 0 \rangle \,
\end{equation}
thus evading HT, in agreement with Appendix~\ref{sub:evasion}.

\section{A finite perturbation theory}\label{SEC:loops}

In this section, we compute the two-point and four-point Green functions in perturbation theory within the DRQFT.

\paragraph{The Feynman propagator.} The Feynman propagator  $\Delta_F(x,y)$ given by
\begin{align}\label{k3propagator}
\Delta_F(x,y) & =  \langle 0 | \mathcal{T} \phi(x)\phi(y) | 0 \rangle\equiv\Theta(x_0-y_0)\langle 0 |  \phi(x)\phi(y)| 0 \rangle +  \Theta(y_0-x_0)\langle 0 |  \phi(y)\phi(x)| 0 \rangle  \nonumber \\\nonumber \\
&= \frac{\Theta(x_0-y_0)}{(2\pi)^3} \left(\int \frac{d\bar{k} \, r(\bar{k} )}{2 \omega_k} e^{-i k (x-y)} \right) + \frac{\Theta(y_0-x_0)}{(2\pi)^3} \left(\int \frac{d\bar{k} \, r(\bar{k} )}{2 \omega_k} e^{i k (x-y)} \right)\,,
\end{align}
where  $\mathcal{T}$  denotes the time-order operator, as defined above. It is straightforward to show that  $\Delta_F(x,y)$ is  the Green function associated with the differential operator in Eq.~\eqref{KGE}, since
\begin{align}\label{feynman}
\left(\Box + \frac{1}{r(x)}\partial_{\mu}r(x)\partial^{\mu}-m^2 \right)\Delta_F(x,y)&= -i \delta(x-y) +\nonumber \\
 & \frac{\partial_{0}r(x)}{r(x)}\delta^{(1)}(x_0-y_0)\langle 0 | \left[ \phi(x),\phi(y)\right]| 0 \rangle \,.
\end{align}
From the equal time canonical commutation relations, one can immediately see that the second term in Eq.~\eqref{feynman} vanishes.
Next, using the residue theorem for evaluating the first term in Eq.~\eqref{feynman}  at the pole $k_0=\omega_k$ gives
\begin{equation}
\frac{i}{(2\pi)^4} \int \frac{dk \, r(k) e^{-i k (x-y)}}{(k_0-\omega_k)(k_0+\omega_k)} = \frac{i}{(2\pi)^4} \int \frac{dk_0 \, d\bar{k}     \, r(k) e^{-i k (x-y)}}{(k_0-\omega_k)(k_0+\omega_k)} = \frac{1 }{(2\pi)^3} \int \frac{d\bar{k} \, r(\bar{k} ) e^{-i k (x-y)}}{2 \omega_k}\,.
\end{equation}
A similar calculation applies for the second term in Eq.~\eqref{k3propagator}, and by the standard Feynman prescription on contour integration the Eq.~\eqref{k3propagator} becomes
\begin{equation}\label{k4propagator}
\Delta_F(x,y) = \frac{i}{(2\pi)^4} \int  \frac{dk \, r(k)}{k^2-m^2 +i \epsilon}\, e^{- i \,p\cdot  (x-y)}\,.
\end{equation}

\paragraph{One-loop corrections.} We now consider the one-loop corrections, induced by the interaction term in the Lagrangian
\begin{equation}
\mathcal{L}_{int}=- \frac{\lambda}{4!}\phi^4\,.
\end{equation}
From Eq.~\eqref{k4propagator}, one expects loop calculations standard-like but modified due to the function $r$. Indeed, it can be shown that in a Feynman diagram with $I$ internal lines, $E$ external lines, and $L$ number of legs in the vertex, the function  $r$  enters in the integration of virtual momenta with the power $N=\left(\frac{L-2}{L}\right)I-E/L+1$, which mean one $r$ for each loop integration $k$. In particular,
the one-loop two-point correlator reads
\begin{equation}\label{masscorrect}
\Gamma_{\text{one-loop}}^{(2)}(p_1,p_2) \propto (2\pi)^4\,\delta(p_1+p_2)\int dk \, r(k) \frac{1}{k^2-m^2}\,.
\end{equation}

Similarly, the one-loop  4-point Green function  is given by
\begin{equation}\label{G4c}
\Gamma^{(4)}_{\text{one-loop}}(p_1, p_2, p_3,p_4) = const  \times (2\pi)^4\, \delta(s+q) \int dk \, r(k)\Delta_F (s-k) \Delta_F (k)\,,
\end{equation}
and $s=p_1+p_2$ and $q=p_3+p_4$.

For the most divergent integral of Eq.~\eqref{masscorrect} to be finite, $r(k)$ must scale at least as $|k|^{-\eta}$, with $\eta>2$. For the sake of illustration, in what follows, we consider $\eta=3$:
\begin{equation}\label{R2}
r(k)=\left(\frac{l^2}{l^2+k_\mu l^\mu}\right)^3\,,
\end{equation}
being $l^\mu$ a four-vector parameter with $|l|=M$. Eq.~\eqref{R2} is such that it takes into account the constraint in Eq.~\eqref{constraint}. Moreover, the function $r(k)$ tends to one at long distances, in agreement with standard QFT, which one must recover at low energies. Notice that, although the qualitative behavior of $r(k)$ is determined from the requirement of a four-dimensional space-time at low energy, the specific form to obtain finite results depends on the particular model, which in this case is $\phi(x)^4$ scalar model.

It is worth emphasizing that the choice for the weight function $r(k)$ in Ref.~\cite{Shirkov:2010sh} is not compatible with our constraint.  Notwithstanding these technical differences, the qualitative conclusions found in Ref.~\cite{Shirkov:2010sh} also hold in the DRQFT.

\paragraph{Loop finiteness and low-energy limit. }
Eq.~\eqref{masscorrect} is finite and, for $M\gg m$, it  can be written as
\begin{equation}\label{nore_mass}
\Gamma_{\text{one-loop}}^{(2)}  =  \frac{\lambda  \left(m^2 \log \left(\frac{m}{M}\right)+m^2+M^2\right)}{16 \pi ^2}   \,.
\end{equation}
We now evaluate Eq.~\eqref{G4c} in two limits, namely $p^2\ll M^2$ and $p^2\gg M^2$. The limit $p^2\ll M^2$ gives
\begin{equation}\label{nore_asympt_1}
\Gamma_{\text{one-loop}}^{(4)}  =   \frac{3 \lambda ^2 \log\left(\frac{p^2}{4 M^2}\right)+1}{32 \pi ^2}+ \mathcal{O}(p^2/M^2) \,,
\end{equation}
while for  $p^2\gg M^2$ we obtain
\begin{equation}\label{nore_asympt_2}
\Gamma_{\text{one-loop}}^{(4)}=   \frac{3 \lambda ^2 M^2 \left(4 M^2 \log \left(\frac{p^2}{4 M^2}\right)-4 M^2+p^2\right)}{8 \pi^2 p^4} + \mathcal{O}(M^6/p^6)  \,.
\end{equation}
Since the Green functions obey the Callan-Symanzik equation, just as in the standard QFT -- also recall Sec.~\ref{sec:motivation} -- one is free to implement an arbitrary, finite subtraction to (re)normalize Eq.~\eqref{nore_asympt_1} to
\begin{equation}\label{asympt_1}
\lambda_R^{p\ll M}(p) \simeq \lambda+\frac{3 \lambda^2 \log \left(\frac{p^2}{\mu_0^2}\right)}{32\pi ^2}\,,
\end{equation}
where $\lambda=\lambda(\mu_0)$.

The same subtraction applied to Eq.~\eqref{nore_asympt_2} gives in the limit $p^2\gg M^2$
\begin{equation}\label{asympt_2}
\lambda_R^{p\gg M}(p) \simeq \lambda +\frac{3 \lambda ^2 \left(\log \left(\frac{4 M^2}{\mu_0^2}\right)+\frac{4 M^2 \left(4 M^2\log \left(\frac{p^2}{4 M^2}\right)-4 M^2+p^2\right)}{p^4}-1\right)}{32 \pi ^2}\,.
\end{equation}
The Eq.~\eqref{asympt_1} must be compared with the one-loop running in the standard QFT
\begin{equation}\label{standard2}
\lambda_R^{standard}(\mu)= \frac{\lambda}{1-\frac{\beta_1}{2}\lambda \log\left(\frac{\mu^2}{\mu_0^2}\right)} \simeq \lambda+ \frac{\beta_1}{2}\lambda^2 \log\left(\frac{\mu^2}{\mu_0^2}\right)\,,
\end{equation}
with $\beta_1=\frac{3}{16\pi^2}$, which matches to Eq.~\eqref{asympt_1} for $\mu^2=p^2$. Hence the DRQFT reproduces the standard QFT result in the low energy limit $p^2\ll M^2$. In other words, it reproduces -- in this limit -- the usual renormalization, based on dimensional regularization. 

On the other hand, in the limit $p\gg M$, the coupling $\lambda_R$ (in Eq.~\eqref{asympt_2})
rapidly approaches a constant value. Therefore, the theory has an asymptotic UV fixed point at the one-loop level. Since the higher loop corrections are automatically finite, it is guaranteed that higher order terms are subleading and hence do not alter the qualitative behavior of Eq.~\eqref{asympt_2}.
Moreover, the absence of UV renormalons -- which we shall discuss in the next subsection -- implies no incalculable large-order contributions, and the coupling $\lambda$ remains small at all energies (if small at low energies).

Finally, in terms of the hard conjunction of Ref.~\cite{Shirkov:2010sh}, the running of $\lambda$ in Eqs.~\eqref{asympt_1} and~\eqref{asympt_2} corresponds to the running of a 4D QFT at low energies and a 1D QFT at higher energies. Recall that for the function $r$ in Eq.~\eqref{R2} to give finite result in Eq.~\eqref{masscorrect}, it is sufficient for its exponent  $\eta$  to be larger than 2 -- although we have fixed it, for simplicity, equal to 3. In turn, the requirement of finite loop corrections implies that the reduced dimension must be smaller than two, in agreement with the insight from the quantum gravity approach suggesting that the space-time dimension could be smaller, but close to two ~\cite{Ambjorn:2005db}.

\subsection{Absence of renormalon singularities}

We now show that renormalon singularities are absent in DRQFT. For clarity, we first recall how renormalons appear in standard QFT. In the $\phi^4$ model, a specific realization of the UV renormalons can be derived from the 't Hooft's skeleton diagram~\cite{tHooft:1977xjm}. Denoting it with $S_n$, with $n$-bubbles, one has
\begin{align}\label{skeleton}
S_n =& \,\, \includegraphics[scale=0.7]{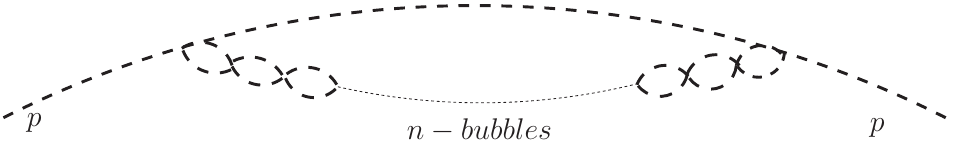} \nonumber  \\
& \nonumber \\
& \propto \lambda^{n+1} \int dk \frac{1}{(k-p)^2-m^2} \, B(k)^n \,,
\end{align}
where $B(k)$ denotes the one-loop correction of the four-point function in the $\phi^4$ model. Notice that from Eq.~\eqref{standard2} and for large momentum $k$, $B(k)$ is proportional to $\beta_1 \log(k/\mu)$. Expanding for large $k$ Eq.~\eqref{skeleton}, replacing $B(k)\propto \beta_1 \log(k/\mu)$, and reabsorbing the
divergent part of Eq.\eqref{skeleton} in the proper counterterm, one gets the $n!$ behavior
\begin{equation}\label{factorial}
S_n \approx \lambda^{n+1} \beta_1^n n! \,.
\end{equation}
The  Borel transform ($\mathcal{B}: \lambda\mapsto z $)  for the above equation is given by
\begin{equation}
\mathcal{B}\left[S_n\right] =  \frac{1}{1-z \, \beta_1}\,.
\end{equation}
The pole at $z=1/\beta_1$ is the first UV renormalon. Considering higher orders in $k$ in the expansion of Eq.~\eqref{skeleton}, one gets additional renormalon singularities at $z=u/\beta_1$ ($u=1,2,3,...$). These latter singularities imply an infinite number of ambiguities in the Laplace integral, making it thus ill-defined and hampering the Borel-Laplace resummation in the scalar model.\\

Consider now Eq.~\eqref{skeleton} from the point of view of the DRQFT. Evident modifications concerning the standard case are  $dk\rightarrow dk  \, r(k)$ and, more importantly, the function $B(k)$ approaches a constant value for large momentum $k$, as can be seen from Eq.~\eqref{asympt_2}. Consequently, there is no  $\log^n$ contribution in the integral of Eq.~\eqref{skeleton}, which is the source of the $n!$ contribution in Eq.~\eqref{factorial}. Therefore, there are no UV renormalons for the $\phi^4$ model within DRQFT.

Note that both HT is evaded, and no renormalon ambiguities appear. This agrees with the conjecture of Ref.~\cite{Maiezza:2020qib}, where it is proposed that renormalons could be understood as a consequence of HT.

\subsection{On the analyticity domain of the two-point function}

Renormalons are not the only source of problems when making sense out of QFT in four dimensions, and other problems also arise, such as the superexponential behavior of the Borel transform~\cite{tHooft:1977xjm}~\footnote{Notice that there are diagrams -- pointed out inRef.~\cite{deCalan:1981szv} -- which although do not give poles in the Borel transform at finite locations, they destroy the Borel-Laplace re-summability. This means that these contributions, called ``renormalons at infinity", make the Laplace integral not convergent, albeit Borel transform ``locally exists".
Therefore, we interpret these ``renormalons at infinity" as the diagrammatic counterpart of the argument for the superexponential behavior found in Ref.~\cite{tHooft:1977xjm}.}.
The superexponential behavior is derived from the two-point Green function's accumulation of singularities for infinitesimal coupling. The latter argument was built for pure Yang-Mills (so asymptotically free) models since there are no UV renormalons on the positive semi-axis for this model. Hence, it makes sense to investigate the presence of additional inconsistencies. In the same vein, since we are also arguing that there are no UV renormalons on the positive axis in the DRQFT, even for the $\phi^4$ model, it is worth asking whether the same problems arise. For clarity, we first review the original argument, which relies on two footholds: the non-perturbative insight from the Kallen-Lehmann representation of the Green function and the one-loop running for the coupling $\lambda$.

The Green function singularities~\footnote{The singularities for Green functions can be derived from the Kallen-Lehmann spectral representation.} for Minkowskian momentum are all located in the positive real axis, where in addition to simple poles for the one-particle state, there is a branch-cut starting at the multiparticle energy threshold $p>2 m$. The idea is to understand the implications of this branch cut in momentum space for the coupling constant dependence of the Green function. Following Ref.~\cite{tHooft:1977xjm}, the Green function at one-loop order can be seen as a function of the single variable $ X:=\frac{1}{\lambda_R(k^2)}+\frac{\beta_{1}}{2} \log \left(-k^{2} / \mu_{0}^{2}\right)$, namely,
\begin{equation}\label{Gcoupling}
G^{(2)}(X) =  G^{(2)}\left( \frac{1}{\lambda_R(k^2)}+\frac{\beta_{1}}{2} \log \left(-k^{2} / \mu_{0}^{2}\right)\right)\,,
\end{equation}
where the standard one-loop running shown in Eq.~\ref{standard2} is used. The next step is to study the analytic structure of the above Green function for complex $X$. The variable $X$ can be complex for real coupling and complex momentum and it can also be complex for real momentum and complex coupling. For complex $k^2$ and real coupling $\lambda_R$, the Green function has singularities when $k^2$ is real and positive (Minkowskian), i.e.
\begin{equation}
\frac{1}{\lambda_R(k^2)}+\frac{\beta_{1}}{2} \log \left(-k^{2} / \mu_{0}^{2}\right) = \frac{1}{\lambda_R(k^2)}+\frac{\beta_{1}}{2} \log \left(k^{2} / \mu_{0}^{2}\right) +  \frac{\beta_1}{2}(2n+1) \pi i = R +\frac{\beta_1}{2}(2n+1) \pi i   \,,
\end{equation}
and $R$ is any real number and $n$ is a natural number.

\begin{figure}[t]
\centerline{
\includegraphics[width=0.5\columnwidth]{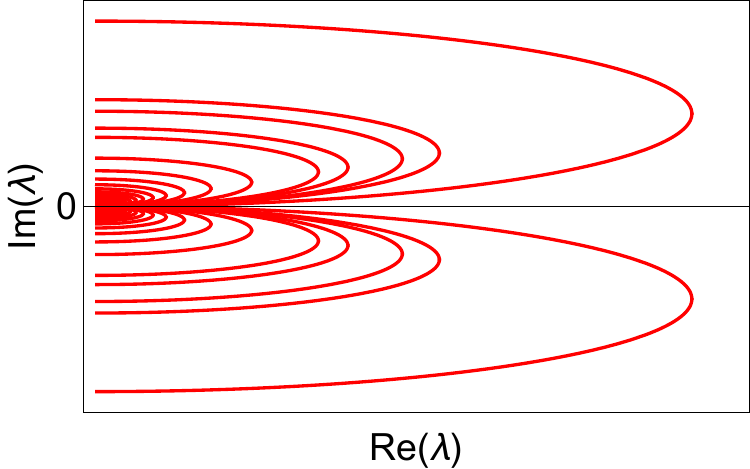}%
}
\caption{Horned shaped analyticity domain of the two-point correlator in standard quantum field theory derived from Eq.~\eqref{thooft_sing}.
Red lines denote the singularities for complex values of the coupling $\lambda$. Notice the accumulation of singularities around the origin.}
\label{figanalytic}
\end{figure}

Conversely, one can also analyze the case of real momentum and complex coupling. The crucial point is that for real Euclidean momentum $k^2<0$, the $\log$ term in Eq.~\eqref{Gcoupling} will not reproduce the known singularities known from the Kallen-Lehmann representation, and it may seem that one can avoid the Kallen-Lehmann singularity in the Euclidean region. This is not possible, and the reason is that the Green function is a function of $X$ only. Therefore, it must be that for Euclidean momentum; the Kallen-Lehmann singularity  manifests in the momentum dependence of the complex coupling  $\lambda(k^2)$ as follows
\begin{equation}\label{thooft_sing}
\frac{1}{(\lambda_R(k^2))_{sing}} =   R +  \frac{\beta_1}{2}(2n+1) \pi i  \,.
\end{equation}
In the complex $\lambda_R$ plane, the Eq.~\eqref{thooft_sing} leads to the ``horned shaped" domain~\footnote{t'Hooft argument was also reproduced in the formalism of resurgence (accelero-summation)~\cite{Bellon:2018lwy}.}  shown in Fig.~\ref{figanalytic}. The lines represent the singularities on the complex plane. Notice the accumulation of singularities at the origin. Such a bad analytic behavior precludes any analytic continuation to finite values of the coupling constant and, in particular, implies a superexponential behavior for the Borel transform of the Green functions~\cite{tHooft:1977xjm}.\\

In analogy with the renormalon issue, we now argue that the problem sketched above and visualized in Fig.~\ref{figanalytic} is absent in the dimensionally-reduced QFT.

For this purpose, notice that Eq.~\eqref{thooft_sing} follows by considering the renormalized, running coupling in the deep UV in Eq.~\eqref{standard2}. In particular, the piece $\propto (2 n +1)\pi i$ comes from the multi-valued $\log$ function of the standard running in Eq.~\eqref{standard2}. The presence of an \emph{unsuppressed} imaginary piece is essential to obtain the analytic structure shown in Fig.~\ref{figanalytic}. In the deep ultraviolet limit $p^2\gg M^2$ of  DRQFT, the running is given by Eq.~\eqref{asympt_2}, where the $\log$ piece is suppressed, in contrast to  Eq.~\eqref{standard2}. In particular, the imaginary part in Eq.~\eqref{asympt_2} goes as
\begin{equation}
\frac{1}{(\lambda_R^{p>>M})_{sing}} \approx R +\frac{3  M^4}{2 \pi ^2  p^4}(2n+1)\pi i \,.
\end{equation}
and thus, the Kallen-Lehmann singularities for complex $\lambda_R$ are suppressed at high momenta. In this case, there is an isolated singularity at $\lambda_R=0$ in the limit $p\rightarrow \infty$.

As first noted in Ref.~\cite{Dyson:1952tj}, there is a connection between such a singularity when $\lambda_R=0$ and the divergent asymptotic expansion in the coupling. This suggests that the isolated singularity is linked to the instanton's large-order $n!$ contributions, and the semi-classical nature of the instantons renders them conceptually harmless. Moreover, recent resurgent techniques allow us to tackle them -- for example, see~\cite{Basar:2013eka,Dunne:2013ada,Dorigoni:2014hea,Borinsky:2022knn}. Therefore, we conclude that the analytic structure of the Green functions in the coupling complex plane is such that no insurmountable obstacles appear for an \emph{exact} non-perturbative renormalization within DRQFT.

\section{Outlook}\label{Discussion}

Importing the notion of dimensional reduction in quantum gravity to QFT can lead to a formally consistent and finite theory at all energies. While attempts to formulate a finite QFT are not new~\cite{Altaisky:2006dj,Altaisky:2010wv}, our proposal goes beyond the current literature in several points: we study the interplay between Haag's theorem, the Feynman diagrams finitude, and in particular, their finitude beyond perturbation theory. We assess the latter point through the Borel resummability, the standard way to analytically continue quantum field theory from the perturbative to the non-perturbative regime. We show that the renormalon issue and superexponential behavior of the Green function as a function of the coupling are absent in the dimensional reduced QFT.

We have illustrated our results in a scalar field model, and the generalization to gauge models may have additional subtleties. In particular,  one may wonder whether a dimensional scale $M$ conflicts with the gauge symmetry principle, in analogy with the cutoff regularization. While providing an answer to the latter question is beyond the scope of this work, we limit ourselves to give a heuristic argument suggesting the consistency of DRQFT with the gauge principle. It is known that a sharp cutoff is not compatible with gauge invariance since  a gauge transformation reads in momentum space as $p_\mu \rightarrow p_\mu - i e A_\mu(p)$ (being $A_\mu$ the gauge field. Demanding $|p|<$ cutoff is equivalent to forbidding gauge transformations for modes above the cutoff.
In contrast, DRQFT keeps all the modes, just as dimensional regularization. The energy scale $M$ does not act as a cutoff, and DRQFT might be applicable in gauge theory.  

However, the subject requires a dedicated analysis, and, were the answer positive, it would open up the possibility of applications to the standard model. In this case, one would have a change in the running of its parameters. One expects that the running would reach asymptotically constant values above the scale $M$, making the model consistent with asymptotically safe quantum gravity~\cite{weinberg1979ultraviolet}, which, in turn, may be intimately related to dimensional reduction~\cite{Niedermaier:2003fz,Lauscher:2005qz}.

One further speculation is on a possible consequence of the DRQFT on the Higgs mass hierarchy problem. This is usually stated in terms of hypothetical corrections to its mass that are proportional to a new heavy energy scale $\Lambda$. In renormalized perturbation theory, the problem sounds immaterial since the mass' UV dependence can be eliminated ``at any order" by the appropriate renormalization conditions. The renormalization conditions are such that the high scale disappears from the \emph{renormalized} Lagrangian.  Notice that  ``at any order" means that loop corrections can be considered up $\lambda^n$, with $n$ arbitrarily large but finite. Since renormalization is not a convergent and complete procedure -- at least spoiled by the presence of renormalon singularities -- the hierarchy problem can become real beyond perturbation theory. In other words, the issue of the Higgs mass is a non-perturbative one, i.e., in the limit $n\rightarrow\infty$. This is a crucial point, sometimes missed in the literature. Recently, the authors of Ref.~\cite{Maiezza:2020nbe}, based on the resurgent approach of Refs.~\cite{Maiezza:2019dht,Bersini:2019axn}, showed that indeed the scalar mass receives a correction from the renormalons proportional to the non-perturbative Landau pole scale (for standard QFT). All these difficulties are not present in the DRQFT since there are no renormalons, and the renormalization conditions are well-defined in the limit $n\rightarrow\infty$.

\section*{Acknowledgements}

We thank Oleg Antipin and Jagu Jagannathan for their useful comments on the manuscript, and Emanuela Pichelli for discussions. 

\appendix

\section{Haag's theorem and the interaction picture in quantum field theory}\label{SEC:Haag}

HT states that if the free and interactive fields are related by a unitary matrix (Dyson matrix), then the free and interactive vacua coincide. In turn, this implies that all the correlators of the free and interactive fields are the same: the interaction picture in QFT can be built only in the trivial, non-interactive case. For completeness, here we quote the argument leading to HT in the language of standard QFT as in Ref.~\cite{Maiezza:2020qib}, in contrast with the rigorous proof in the axiomatic formalism~\cite{hall1957theorem,LOPUSZANSKI1962169} (see the review~\cite{Klaczynski:2016qru})

Let us consider the free scalar field $\phi_0$ acting in the Hilbert space $\mathcal{H}_0$, and the interactive scalar field $\phi$ acting in the Hilbert space $\mathcal{H}$.

\paragraph{Free and interactive fields are Poincar\'e covariant.} The spatial \emph{continuous} translation operators $T_0 (T)\in \mathcal{H}_0 (\mathcal{H})$, respectively, act on $\phi_0(\phi)$ as,
\begin{align}\label{translation}
  T_0^\dagger \phi_0(x) T_0 =&  \phi_0(x_0,\bar{x}-\bar{b}) \nonumber \\
  T^\dagger \phi(x) T =&  \phi(x_0,\bar{x}-\bar{b})\,,
\end{align}
being $\bar{b}$ a vector parameter associated with translations.

\paragraph{Vacua are Poincar\'e invariant.} The vacua $|0\rangle (|\Omega\rangle) \in \mathcal{H}_0 (\mathcal{H})$, respectively, are translational invariant,
\begin{align}\label{vacua}
 T_0|0\rangle =& |0\rangle  \nonumber \\
 T|\Omega\rangle =& |\Omega\rangle\,.
\end{align}
\paragraph{Unitary equivalence of the free and interacting fields. }In the interaction picture, $\phi_0$ and $\phi$ are related by the Dyson unitary matrix $U$:
\begin{equation}\label{unitary}
\phi = U^\dagger \phi_0 U\,.
\end{equation}
Combining Eqs.~\eqref{translation} and~\eqref{unitary}, the following chain of equalities holds:
\begin{equation}\label{chain}
\phi(x_0,\bar{x}-\bar{b})=T^\dagger \phi(x) T = T^\dagger U^\dagger \phi_0(x) U T = U^\dagger \phi_0(x_0,\bar{x}-\bar{b}) U = U^\dagger T_0^\dagger \phi_0(x) T_0 U\,,
\end{equation}
which implies
\begin{equation}\label{UT}
U T = T_0 U \,.
\end{equation}
Multiplying on the right Eq.~\eqref{UT} for $|\Omega\rangle$ and using Eq.~\eqref{vacua}, one obtains
\begin{align}
  U T |\Omega\rangle  =& T_0 U |\Omega\rangle \nonumber \\
  U |\Omega\rangle =& T_0 U |\Omega\rangle\,,
\end{align}
and employing again Eq.~\eqref{vacua}, one finds from the latter
\begin{equation}\label{I}
U |\Omega\rangle = |0\rangle
\end{equation}
or
\begin{equation}\label{II}
|\Omega\rangle = U^\dagger |0\rangle\,.
\end{equation}
Finally, multiplying Eq.~\eqref{I} by $|\Omega\rangle$ (on the left), and Eq.~\eqref{II} by $|0\rangle$ (on the left) yields
\begin{equation}
\langle \Omega | U | \Omega \rangle = \langle 0 | U | 0 \rangle\,,
\end{equation}
which implies
\begin{equation}\label{thesis}
| \Omega \rangle = | 0 \rangle\,.
\end{equation}
This shows that the free and interactive vacua coincide. In principle, this prevents the construction of the Gell-Mann and Low formula that is the basis of any amplitude calculations. The latter implies that the S-matrix is of the form
\begin{equation}
S=\lim_{t\rightarrow \infty} U(-t,t)=\mathds{1} \,,
\end{equation}
being $t$ the time. Therefore, within the interaction picture in standard QFT, the matrix $S$ exists only in the trivial case of free fields.

\section{Evading the no-go imposed by Haag's theorem}\label{sub:evasion}

Let us start by enumerating the basic assumptions used above, on which HT relies:
1) the fields are Poincar\'e covariant, in particular, \emph{continuous} translation symmetry transformations act on the fields as in Eq.~\eqref{translation}; 2) the quantum vacuum state $ | 0 \rangle$ is Poincar\'e invariant; 3) the free and the interacting fields are unitary equivalent, i.e., they are related via Dyson's matrix as in Eq.~\eqref{unitary}.

For example, if defined on a discrete space-time lattice, QFT avoids HT because it breaks assumption 1) while maintaining assumptions 2) and 3). In this work, we consider the possibility of evading assumption 2) while keeping 1) and 3).

One way to break assumption 2) is by assuming that
the translation operator $T_0$ in Eq.~\eqref{translation} is time-dependent $T_0=T_0(t)$. More precisely, the time dependence of $T_0$ is due to a time-dependent momentum operator, such that
\begin{equation}\label{absurdum}
T_0(t) |0\rangle =e^{-i \vec{P}(t)\cdot  \vec{x}} |0\rangle  =|0\rangle  \,\,\,\text{or} \,\,\, \langle 0| T_0^\dagger(t) = \langle 0| \,\,\,\,\,\,\, \forall t \,,
\end{equation}
where $P$ denotes the canonical momentum of the field shown in Eq.~\eqref{3momentum}. One can then consider $T_0^\dagger(t) T_0(t+\Delta t)\neq 1$, and to first order in $\vec{P}$ one can show that
\begin{equation}\label{TT}
T_0^\dagger(t) T_0(t+\Delta t)=1-i \dot{\vec{P}}\cdot \vec{x} \Delta t + \mathcal{O}(\Delta t)^2\,,
\end{equation}
with $\dot{\vec{P}}$  given in Eq.~\eqref{pdot}.

Assuming the vacuum is invariant under translation in space for any given time $t$, one can see that using Eq.~\eqref{absurdum}, the following equality holds
\begin{equation}\label{case1}
\langle 0 | T_0^\dagger(t) T_0(t+\Delta t) |0\rangle = \langle 0|0\rangle=1\,
\end{equation}
while using Eq.~\eqref{TT} gives
\begin{equation}\label{case2}
\langle 0 | T_0^\dagger(t) T_0(t+\Delta t) |0\rangle = 1-i\Delta t \langle 0 | \dot{\vec{P}}|0\rangle\cdot \vec{x}  + \mathcal{O}(\Delta t)^2\,.
\end{equation}
If $ \langle 0 | \dot{\vec{P}}|0\rangle\neq0$ -- it is the case of the DRQFT -- one reaches a \emph{contradiction} between Eqs.~\eqref{case1} and~\eqref{case2}, and thus must be that
\begin{equation}\label{inequality}
T_0(t) |0\rangle \neq |0\rangle\,.
\end{equation}
The above equation invalidates Eqs.~\eqref{I} and~\eqref{II}, and it is thus an explicit realization of a quantum vacuum that is not invariant under space translations.

The generic mechanism to bypass HT, presented in this appendix, is achieved in the DRQFT because the non-standard integration measure introduces time dependence into the translational operator. In particular, Eq.~\eqref{achievement_evasion} matches Eq.~\eqref{inequality}.

\bibliographystyle{jhep}
\bibliography{biblio}

\providecommand{\href}[2]{#2}\begingroup\raggedright\begin{thebibliography}{10}

\bibitem{Haag1964}
R.~Haag and D.~Kastler, \emph{An algebraic approach to quantum field theory},
  \href{http://dx.doi.org/10.1063/1.1704187}{\emph{Journal of Mathematical
  Physics} {\bfseries 5} (July, 1964) 848--861}.

\bibitem{PhysRev.112.669}
R.~Haag, \emph{Quantum field theories with composite particles and asymptotic
  conditions}, \href{http://dx.doi.org/10.1103/PhysRev.112.669}{\emph{Phys.
  Rev.} {\bfseries 112} (Oct, 1958) 669--673}.

\bibitem{ruelle1962asymptotic}
D.~Ruelle, \emph{On asymptotic condition in quantum field theory},
  {\emph{Helvetica Physica Acta} {\bfseries 35} (1962) 147}.

\bibitem{Haag:1955ev}
R.~Haag, \emph{{On quantum field theories}}, {\emph{Kong. Dan. Vid. Sel. Mat.
  Fys. Med.} {\bfseries 29N12} (1955) 1--37}.

\bibitem{tHooft:1977xjm}
G.~'t~Hooft, \emph{{Can We Make Sense Out of Quantum Chromodynamics?}},
  {\emph{Subnucl. Ser.} {\bfseries 15} (1979) 943}.

\bibitem{deCalan:1981szv}
C.~de~Calan and V.~Rivasseau, \emph{{Local Existence of the Borel Transform in
  Euclidean $\phi^4$ in Four-dimensions}},
  \href{http://dx.doi.org/10.1007/BF01206946}{\emph{Commun. Math. Phys.}
  {\bfseries 82} (1981) 69}.

\bibitem{glimm2012quantum}
J.~Glimm and A.~Jaffe, \emph{Quantum physics: a functional integral point of
  view}.
\newblock Springer Science \& Business Media, 2012.

\bibitem{tHooft:1993dmi}
G.~'t~Hooft, \emph{{Dimensional reduction in quantum gravity}}, {\emph{Conf.
  Proc. C} {\bfseries 930308} (1993) 284--296},
  [\href{https://arxiv.org/abs/gr-qc/9310026}{{\ttfamily gr-qc/9310026}}].

\bibitem{Lauscher:2005qz}
O.~Lauscher and M.~Reuter, \emph{{Fractal spacetime structure in asymptotically
  safe gravity}},
  \href{http://dx.doi.org/10.1088/1126-6708/2005/10/050}{\emph{JHEP} {\bfseries
  10} (2005) 050}, [\href{https://arxiv.org/abs/hep-th/0508202}{{\ttfamily
  hep-th/0508202}}].

\bibitem{Ambjorn:2005db}
J.~Ambjorn, J.~Jurkiewicz and R.~Loll, \emph{{Spectral dimension of the
  universe}},
  \href{http://dx.doi.org/10.1103/PhysRevLett.95.171301}{\emph{Phys. Rev.
  Lett.} {\bfseries 95} (2005) 171301},
  [\href{https://arxiv.org/abs/hep-th/0505113}{{\ttfamily hep-th/0505113}}].

\bibitem{Benedetti:2008gu}
D.~Benedetti, \emph{{Fractal properties of quantum spacetime}},
  \href{http://dx.doi.org/10.1103/PhysRevLett.102.111303}{\emph{Phys. Rev.
  Lett.} {\bfseries 102} (2009) 111303},
  [\href{https://arxiv.org/abs/0811.1396}{{\ttfamily 0811.1396}}].

\bibitem{Calcagni:2009kc}
G.~Calcagni, \emph{{Fractal universe and quantum gravity}},
  \href{http://dx.doi.org/10.1103/PhysRevLett.104.251301}{\emph{Phys. Rev.
  Lett.} {\bfseries 104} (2010) 251301},
  [\href{https://arxiv.org/abs/0912.3142}{{\ttfamily 0912.3142}}].

\bibitem{Calcagni:2010bj}
G.~Calcagni, \emph{{Quantum field theory, gravity and cosmology in a fractal
  universe}}, \href{http://dx.doi.org/10.1007/JHEP03(2010)120}{\emph{JHEP}
  {\bfseries 03} (2010) 120},
  [\href{https://arxiv.org/abs/1001.0571}{{\ttfamily 1001.0571}}].

\bibitem{Shirkov:2010sh}
D.~V. Shirkov, \emph{{Coupling running through the Looking-Glass of dimensional
  Reduction}}, \href{http://dx.doi.org/10.1134/S1547477110060014}{\emph{Phys.
  Part. Nucl. Lett.} {\bfseries 7} (2010) 379--383},
  [\href{https://arxiv.org/abs/1004.1510}{{\ttfamily 1004.1510}}].

\bibitem{Modesto:2011kw}
L.~Modesto, \emph{{Super-renormalizable Quantum Gravity}},
  \href{http://dx.doi.org/10.1103/PhysRevD.86.044005}{\emph{Phys. Rev. D}
  {\bfseries 86} (2012) 044005},
  [\href{https://arxiv.org/abs/1107.2403}{{\ttfamily 1107.2403}}].

\bibitem{Anchordoqui:2010er}
L.~Anchordoqui, D.~C. Dai, M.~Fairbairn, G.~Landsberg and D.~Stojkovic,
  \emph{{Vanishing Dimensions and Planar Events at the LHC}},
  \href{http://dx.doi.org/10.1142/S0217732312500216}{\emph{Mod. Phys. Lett. A}
  {\bfseries 27} (2012) 1250021},
  [\href{https://arxiv.org/abs/1003.5914}{{\ttfamily 1003.5914}}].

\bibitem{Mureika:2011bv}
J.~R. Mureika and D.~Stojkovic, \emph{{Detecting Vanishing Dimensions Via
  Primordial Gravitational Wave Astronomy}},
  \href{http://dx.doi.org/10.1103/PhysRevLett.106.101101}{\emph{Phys. Rev.
  Lett.} {\bfseries 106} (2011) 101101},
  [\href{https://arxiv.org/abs/1102.3434}{{\ttfamily 1102.3434}}].

\bibitem{Ellis:1992eh}
J.~R. Ellis, N.~E. Mavromatos and D.~V. Nanopoulos, \emph{{String theory
  modifies quantum mechanics}},
  \href{http://dx.doi.org/10.1016/0370-2693(92)91478-R}{\emph{Phys. Lett. B}
  {\bfseries 293} (1992) 37--48},
  [\href{https://arxiv.org/abs/hep-th/9207103}{{\ttfamily hep-th/9207103}}].

\bibitem{Calcagni:2011sz}
G.~Calcagni, \emph{{Geometry and field theory in multi-fractional spacetime}},
  \href{http://dx.doi.org/10.1007/JHEP01(2012)065}{\emph{JHEP} {\bfseries 01}
  (2012) 065}, [\href{https://arxiv.org/abs/1107.5041}{{\ttfamily 1107.5041}}].

\bibitem{svozil1987quantum}
K.~Svozil, \emph{Quantum field theory on fractal spacetime: a new
  regularisation method}, {\emph{Journal of Physics A: Mathematical and
  General} {\bfseries 20} (1987) 3861}.

\bibitem{eyink1989quantum}
G.~Eyink, \emph{Quantum field-theory models on fractal spacetime},
  {\emph{Communications in mathematical physics} {\bfseries 125} (1989)
  613--636}.

\bibitem{Belavin:1975fg}
A.~A. Belavin, A.~M. Polyakov, A.~S. Schwartz and Y.~S. Tyupkin,
  \emph{{Pseudoparticle Solutions of the Yang-Mills Equations}},
  \href{http://dx.doi.org/10.1016/0370-2693(75)90163-X}{\emph{Phys. Lett. B}
  {\bfseries 59} (1975) 85--87}.

\bibitem{tHooft:1976snw}
G.~'t~Hooft, \emph{{Computation of the Quantum Effects Due to a
  Four-Dimensional Pseudoparticle}},
  \href{http://dx.doi.org/10.1103/PhysRevD.14.3432}{\emph{Phys. Rev. D}
  {\bfseries 14} (1976) 3432--3450}.

\bibitem{Coleman:1978ae}
S.~R. Coleman, \emph{{The Uses of Instantons}}, {\emph{Subnucl. Ser.}
  {\bfseries 15} (1979) 805}.

\bibitem{zinn2011barrier}
J.~Zinn-Justin, \emph{Barrier penetration and instantons}, {\emph{Quantum Field
  Theory} {\bfseries 5} (2011) 70}.

\bibitem{Stojkovic:2013xcj}
D.~Stojkovic, \emph{{Vanishing dimensions}: {A review}},
  \href{http://dx.doi.org/10.1142/S0217732313300346}{\emph{Mod. Phys. Lett. A}
  {\bfseries 28} (2013) 1330034},
  [\href{https://arxiv.org/abs/1406.2696}{{\ttfamily 1406.2696}}].

\bibitem{Carlip:2017eud}
S.~Carlip, \emph{{Dimension and Dimensional Reduction in Quantum Gravity}},
  \href{http://dx.doi.org/10.1088/1361-6382/aa8535}{\emph{Class. Quant. Grav.}
  {\bfseries 34} (2017) 193001},
  [\href{https://arxiv.org/abs/1705.05417}{{\ttfamily 1705.05417}}].

\bibitem{Lovelace:1983yv}
C.~Lovelace, \emph{{Strings in Curved Space}},
  \href{http://dx.doi.org/10.1016/0370-2693(84)90456-8}{\emph{Phys. Lett. B}
  {\bfseries 135} (1984) 75--77}.

\bibitem{Fradkin:1984pq}
E.~S. Fradkin and A.~A. Tseytlin, \emph{{Effective Field Theory from Quantized
  Strings}}, \href{http://dx.doi.org/10.1016/0370-2693(85)91190-6}{\emph{Phys.
  Lett. B} {\bfseries 158} (1985) 316--322}.

\bibitem{Fradkin:1985fq}
E.~S. Fradkin and A.~A. Tseytlin, \emph{{Effective Action Approach to
  Superstring Theory}},
  \href{http://dx.doi.org/10.1016/0370-2693(85)91468-6}{\emph{Phys. Lett. B}
  {\bfseries 160} (1985) 69--76}.

\bibitem{Callan:1985ia}
C.~G. Callan, Jr., E.~J. Martinec, M.~J. Perry and D.~Friedan, \emph{{Strings
  in Background Fields}},
  \href{http://dx.doi.org/10.1016/0550-3213(85)90506-1}{\emph{Nucl. Phys. B}
  {\bfseries 262} (1985) 593--609}.

\bibitem{PhysRevD.2.1541}
C.~G. Callan, \emph{Broken scale invariance in scalar field theory},
  \href{http://dx.doi.org/10.1103/PhysRevD.2.1541}{\emph{Phys. Rev. D}
  {\bfseries 2} (Oct, 1970) 1541--1547}.

\bibitem{Symanzik:1973pp}
K.~Symanzik, \emph{{Small-distance behaviour in field theory}},
  \href{http://dx.doi.org/10.1007/3-540-07022-2_10}{\emph{Lect. Notes Phys.}
  {\bfseries 32} (1975) 20--72}.

\bibitem{Maiezza:2020qib}
A.~Maiezza and J.~C. Vasquez, \emph{{On Haag\textquoteright{}s Theorem and
  Renormalization Ambiguities}},
  \href{http://dx.doi.org/10.1007/s10701-021-00484-3}{\emph{Found. Phys.}
  {\bfseries 51} (2021) 80},
  [\href{https://arxiv.org/abs/2011.08875}{{\ttfamily 2011.08875}}].

\bibitem{Bellon:2018lwy}
M.~P. Bellon and P.~J. Clavier, \emph{{Analyticity domain of a Quantum Field
  Theory and Accelero-summation}},
  \href{http://dx.doi.org/10.1007/s11005-019-01172-0}{\emph{Lett. Math. Phys.}
  {\bfseries 109} (2019) 2003--2011},
  [\href{https://arxiv.org/abs/1806.08254}{{\ttfamily 1806.08254}}].

\bibitem{Dyson:1952tj}
F.~J. Dyson, \emph{{Divergence of perturbation theory in quantum
  electrodynamics}},
  \href{http://dx.doi.org/10.1103/PhysRev.85.631}{\emph{Phys. Rev.} {\bfseries
  85} (1952) 631--632}.

\bibitem{Basar:2013eka}
G.~Basar, G.~V. Dunne and M.~Unsal, \emph{{Resurgence theory, ghost-instantons,
  and analytic continuation of path integrals}},
  \href{http://dx.doi.org/10.1007/JHEP10(2013)041}{\emph{JHEP} {\bfseries 10}
  (2013) 041}, [\href{https://arxiv.org/abs/1308.1108}{{\ttfamily 1308.1108}}].

\bibitem{Dunne:2013ada}
G.~V. Dunne and M.~\"Unsal, \emph{{Generating nonperturbative physics from
  perturbation theory}},
  \href{http://dx.doi.org/10.1103/PhysRevD.89.041701}{\emph{Phys. Rev. D}
  {\bfseries 89} (2014) 041701},
  [\href{https://arxiv.org/abs/1306.4405}{{\ttfamily 1306.4405}}].

\bibitem{Dorigoni:2014hea}
D.~Dorigoni, \emph{{An Introduction to Resurgence, Trans-Series and Alien
  Calculus}}, \href{http://dx.doi.org/10.1016/j.aop.2019.167914}{\emph{Annals
  Phys.} {\bfseries 409} (2019) 167914},
  [\href{https://arxiv.org/abs/1411.3585}{{\ttfamily 1411.3585}}].

\bibitem{Borinsky:2022knn}
M.~Borinsky and D.~Broadhurst, \emph{{Resonant resurgent asymptotics from
  quantum field theory}},
  \href{http://dx.doi.org/10.1016/j.nuclphysb.2022.115861}{\emph{Nucl. Phys. B}
  {\bfseries 981} (2022) 115861},
  [\href{https://arxiv.org/abs/2202.01513}{{\ttfamily 2202.01513}}].

\bibitem{Altaisky:2006dj}
M.~V. Altaisky, \emph{{Scale-Dependent Functions, Stochastic Quantization and
  Renormalization}}, {\emph{SIGMA} {\bfseries 2} (2006) 046},
  [\href{https://arxiv.org/abs/hep-th/0604170}{{\ttfamily hep-th/0604170}}].

\bibitem{Altaisky:2010wv}
M.~V. Altaisky, \emph{{Quantum field theory without divergences}},
  \href{http://dx.doi.org/10.1103/PhysRevD.81.125003}{\emph{Phys. Rev. D}
  {\bfseries 81} (2010) 125003},
  [\href{https://arxiv.org/abs/1002.2566}{{\ttfamily 1002.2566}}].

\bibitem{weinberg1979ultraviolet}
S.~Weinberg, \emph{Ultraviolet divergences in quantum theories of gravitation},
   in \emph{General relativity}.
\newblock 1979.

\bibitem{Niedermaier:2003fz}
M.~Niedermaier, \emph{{Dimensionally reduced gravity theories are
  asymptotically safe}},
  \href{http://dx.doi.org/10.1016/j.nuclphysb.2003.09.015}{\emph{Nucl. Phys. B}
  {\bfseries 673} (2003) 131--169},
  [\href{https://arxiv.org/abs/hep-th/0304117}{{\ttfamily hep-th/0304117}}].

\bibitem{Maiezza:2020nbe}
A.~Maiezza and J.~C. Vasquez, \emph{{Non-Wilsonian ultraviolet completion via
  transseries}}, \href{http://dx.doi.org/10.1142/S0217751X21500160}{\emph{Int.
  J. Mod. Phys. A} {\bfseries 36} (2021) 2150016},
  [\href{https://arxiv.org/abs/2007.01270}{{\ttfamily 2007.01270}}].

\bibitem{Maiezza:2019dht}
A.~Maiezza and J.~C. Vasquez, \emph{{Non-local Lagrangians from Renormalons and
  Analyzable Functions}},
  \href{http://dx.doi.org/10.1016/j.aop.2019.04.015}{\emph{Annals Phys.}
  {\bfseries 407} (2019) 78--91},
  [\href{https://arxiv.org/abs/1902.05847}{{\ttfamily 1902.05847}}].

\bibitem{Bersini:2019axn}
J.~Bersini, A.~Maiezza and J.~C. Vasquez, \emph{{Resurgence of the
  Renormalization Group Equation}},
  \href{http://dx.doi.org/10.1016/j.aop.2020.168126}{\emph{Annals Phys.}
  {\bfseries 415} (2020) 168126},
  [\href{https://arxiv.org/abs/1910.14507}{{\ttfamily 1910.14507}}].

\bibitem{hall1957theorem}
D.~Hall and A.~Wightman, \emph{A Theorem on Invariant Analytic Functions with
  Applications to Relativistic Quantum Field Theory}.
\newblock Matematisk-fysiske meddelelser. I kommission hos Munksgaard, 1957.

\bibitem{LOPUSZANSKI1962169}
J.~Lopuszański, \emph{A criterion for the free character of fields ii},
  \href{http://dx.doi.org/https://doi.org/10.1016/0029-5582(62)90384-X}{\emph{Nuclear
  Physics} {\bfseries 39} (1962) 169 -- 173}.

\bibitem{Klaczynski:2016qru}
L.~Klaczynski, \emph{{Haag's theorem in renormalised quantum field theories}}.
\newblock PhD thesis, Humboldt U., Berlin, 2016.
\newblock \href{https://arxiv.org/abs/1602.00662}{{\ttfamily 1602.00662}}.
\newblock 10.18452/17448.

\end{thebibliography}\endgroup

\end{document}